\documentclass[prd,showpacs,preprintnumbers,floatfix,twocolumn]{revtex4-1}
\usepackage{natbib,hyperref}
\usepackage{times}
\usepackage{graphicx}
\usepackage{amsmath}
\usepackage{amssymb}
\usepackage{mathrsfs}
\usepackage{slashed}
\usepackage{xcolor}
\usepackage[normalem]{ulem}

\newcommand{\ca}{\mathcal}

\begin{document}

\title{Flavor-singlet baryons in the graded symmetry approach to partially quenched QCD}

\author{Jonathan M. M. Hall} 
\author{Derek B. Leinweber}

\affiliation{Special Research Centre for the Subatomic Structure of Matter (CSSM), Department of Physics,
University of Adelaide, Adelaide, South Australia 5005, Australia}

\begin{abstract}
Progress in the calculation of the electromagnetic properties of
baryon excitations in lattice QCD presents new challenges in the
determination of sea-quark loop contributions to matrix elements.
A reliable estimation of the sea-quark loop contributions represents a
pressing issue in the accurate comparison of lattice QCD results with
experiment.
In this article, an extension of the graded symmetry approach to
partially quenched QCD is presented, which builds on previous theory
by explicitly including flavor-singlet baryons in its construction.
The formalism takes into account the interactions among both octet and
singlet baryons, octet mesons, and their ghost counterparts; the
latter enables the isolation of the quark-flow disconnected sea-quark
loop contributions.
The introduction of flavor-singlet states enables systematic studies
of the internal structure of $\Lambda$-baryon excitations in lattice
QCD, including the topical $\Lambda(1405)$.
\end{abstract}

\maketitle
\preprint{ADP-15-33/T935}

\section{Introduction}
\label{sec:intro}

\subsection{Quark-loop contributions to matrix elements}

\begin{figure}[t]
\begin{flushright}
\includegraphics[width=0.88\hsize]{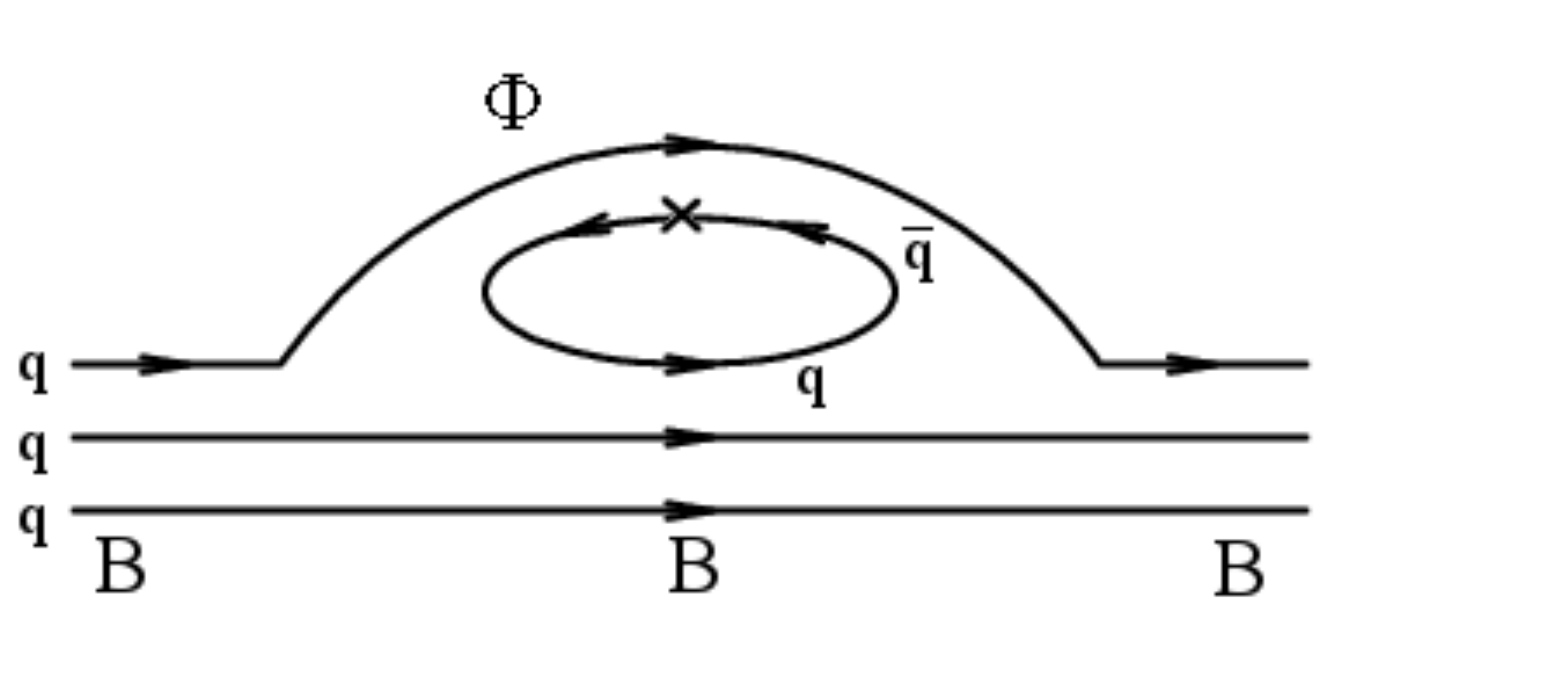}
\end{flushright}
\vspace{-18pt}
\caption{The quark-flow disconnected diagram for the process
  $\mathcal{B}\rightarrow\mathcal{B}\, \Phi$.  The cross on the
  disconnected quark-loop line indicates the insertion of the
  matrix-element current.  For an incoming baryon, $\mathcal{B}$,
  each of the three valence quarks entering on the left hand side may
  contribute to the internal meson state, $\Phi$.  In the graded
  symmetry approach, the conventional $u$, $d$ and $s$ sea-quark loops
  are complemented by their commuting ghost counterparts, $\tilde{u}$,
  $\tilde{d}$ and $\tilde{s}$.
}
\label{fig:Bloop}
\end{figure}

The calculation of hadronic matrix elements in nonperturbative QCD
typically encounters non-trivial contributions from quark-flow
disconnected diagrams in which the current interacts with the
hadrons of interest, first through a virtual sea-quark loop.
An example process for a baryon undergoing a virtual meson-baryon
transition, $\mathcal{B}\rightarrow\mathcal{B}\,\Phi$, involving a
disconnected sea-quark loop, is depicted in Fig.~\ref{fig:Bloop}.
These disconnected-loop diagrams are difficult to calculate in
numerical simulations of QCD due to the space-time coordinate
dependence of both the source and sink of the quark propagator at the
current insertion point.  As the loop is correlated with other quark
degrees of freedom only via gluon exchange, resolving a nontrivial
signal requires high statistics through innovative methods.

While there has been recent success in isolating disconnected-loop
contributions in ground-state baryon matrix elements, even at non-zero
momenta \cite{Green:2015wqa,Sufian:2016pex}, the challenging nature of
these calculations suggests that resolving similar contributions for
hadronic excitations in lattice QCD
\cite{Mahbub:2010rm,Menadue:2011pd,Edwards:2011jj,Mahbub:2012ri,%
  Lang:2012db,Mahbub:2013ala,Morningstar:2013bda,Alexandrou:2014mka,%
  Owen:2015fra,Hall:2013qba,Hall:2014uca,Liu:2015ktc,Leinweber:2015kyz,Liu:2016uzk}
will be elusive.

However, it is here that the meson-baryon dressings of hadron
excitations are particularly important.  Consider, for example, the
$\Lambda(1405)$ resonance which is of particular contemporary
interest.
It is widely agreed that there is a two-pole structure for this
resonance
\cite{Oller2001,GarciaRecio:2002td,Jido:2002yz,Jido:2003cb,Hyodo2004,Magas:2005vu,Geng:2007hz,Ikeda2011,Guo2013,Mai2013,Doring:2010rd,Liu:2016wxq},
associated with attractive interactions in the $\pi\Sigma$ and
$\overline{K}N$ channels.  The two poles lie near the $\pi\Sigma$ and
$\overline{K}N$ thresholds and both channels can contribute to the
flavour singlet components of the resonance, studied in detail herein.
This two-pole structure, as first analyzed in Ref.~\cite{Oller2001},
has now been entered into the Particle Data Group tables \cite{PDG}.
This achievement reflects decades of research using the successful
chiral unitary approaches \cite{Kaiser:1995eg,Kaiser1997,Oset1998,Oset:2001cn,Hyodo:2003jw,Borasoy:2006sr,Ikeda:2011pi,Hyodo:2011qc,Doring:2011xc,Nakamura:2013boa,Sekihara:2013sma,Mai:2014uma,Mai:2014xna}, 
which have
elucidated the two-pole structure \cite{Oller2001}.
In describing the $\Lambda(1405)$ resonance at quark masses larger
than the physical masses, the introduction of a bare state dressed by
meson-baryon interactions has also been explored
\cite{Liu:2016wxq}.  In all cases, the
disconnected loops of Fig.~\ref{fig:Bloop} will make important
contributions to the matrix elements of these states.

Recent lattice QCD calculations of the electromagnetic form factors of
the $\Lambda(1405)$ have reported evidence of an important $\overline
K N$ molecular component in the $\Lambda(1405)$ \cite{Hall:2014uca}.
On the lattice, the strange quark contribution to the magnetic form
factor of the $\Lambda(1405)$ approaches zero, signaling the formation
of a spin-zero kaon in a relative $S$-wave about the nucleon.
However, the light $u$ and $d$ valence-quark sector of the
$\Lambda(1405)$ makes a non-trivial contribution, which can be related
to the proton and neutron electromagnetic form factors.

To understand the physics behind this contribution and enable a
comparison with experiment, one needs to understand the role of
disconnected sea-quark loop contributions to the matrix elements of the
$\Lambda(1405)$.  
These contributions have not been included in
present-day lattice QCD calculations and present a formidable
challenge to the lattice QCD community.

When disconnected sea-quark loop-contributions are unknown, one often
seeks observables insensitive to this contribution.  For example, the
isovector baryon combination is insensitive to a sea-quark
contribution as it is similar for both isospin variants, and cancels.
For baryon states that are not in an isovector formation, a reliable
estimation of the sea-quark loop contributions remains a pressing
issue in the accurate comparison of lattice results with experiment.

\subsection{Flavor-singlet symmetry}

In lattice QCD calculations, the $\Lambda(1405)$ is excited from the
vacuum using interpolating fields dominated by a flavor singlet
operator.  Thus, a complete understanding of the $\Lambda(1405)$
necessarily involves understanding the role of disconnected sea-quark
loop contributions in the flavor-singlet sector of effective field
theory.  Away from the SU(3)-symmetric limit, flavor-singlet and
flavor-octet interpolators mix in isolating the $\Lambda(1405)$
\cite{Menadue:2011pd} and therefore this new information in the
flavor-singlet sector will complement the flavor-octet sector.

Flavor symmetry is important in effective field theory as it provides
guidance on the meson-baryon couplings to be used in coupled-channel
analyses.  It also has the ability to describe the relative strength
between quark-flow connected and disconnected contributions to
observables.  This separation is well understood in the flavor-octet
sector of effective field theory, but is unknown in the
flavor-singlet sector.  This gap in understanding is addressed in
this article.

It is perhaps important to note that the physical $\Lambda(1405)$ is a
mix of flavor symmetries due to the explicit breaking of $SU(3)$
symmetry by the quark masses.  Our task is not to determine this
mixing, but rather to disclose the separation of connected and
disconnected contributions in the flavor-singlet sector for the first
time.  Thus, in referring to flavor-singlet baryons in the following,
it is to be understood that we are referring to the flavor-singlet
component of a baryon.  In this manner, our results are applicable to
$\Lambda$-baryons in general.

\subsection{Graded symmetry approach}

The separation of connected and disconnected contributions to the
baryon couplings can be handled using the graded symmetry approach
\cite{Labrenz:1996jy,Labrenz:1993wt,Savage:2001dy,Chen:2001yi,Beane:2002vq,Detmold:2006vu,Shanahan:2013apa}
or the diagrammatic approach \cite{Leinweber:2002qb,Hall:2013dva}.
Although it has been demonstrated that both methods are consistent
with each other \cite{Leinweber:2002qb,Hall:2013dva}, in this work, the
graded symmetry approach is used because of its ability to be easily
generalized to include flavor-singlet baryons.  The diagrammatic
approach becomes cumbersome when the valence content of the baryon
comprises three unique quark flavors ($u\,d\,s$)
\cite{Leinweber:2002qb}.

Once the coupling strength of disconnected sea-quark loop
contributions to hadron two-point functions is understood, one can use
this information to estimate the size of sea-quark loop contributions
to matrix elements
\cite{Young:2002cj,Leinweber:2003dg,%
Leinweber:2004tc,Leinweber:2006ug,Wang:1900ta,Wang:2013cfp,%
Hall:2013dva}.

The formalism presented herein is based on the original work of
Labrenz and Sharpe \cite{Labrenz:1996jy}, and builds upon the
constructions of that work in a consistent manner.  We create a new, 
completely symmetric baryon field in the graded symmetry Lagrangian, 
and couple this field to octet mesons and octet baryons.  Through the
admission of the commuting `ghost' counterparts, introduced to isolate
the disconnected loop component of each coupling, the particles
collectively fall into representations of the vector subgroup of the
graded symmetry group $SU(3|3)_{V}$, as in
Ref.~\cite{Labrenz:1996jy}.  The anticommuting behavior of the
standard quark degrees of freedom enable our symmetric baryon field in
the graded symmetry Lagrangian to represent the flavor-singlet 
baryon.

Our aim here is to disentangle quark-flow connected and disconnected
couplings between singlet baryons and octet mesons and baryons in
modern $2+1$-flavor lattice QCD simulations.  As such, there are no
issues with the artifacts of quenched QCD, such as a light double-pole
$\eta'$ meson.  Moreover, we will consider quark masses similar to
those of nature, with $m_u \simeq m_d \le m_s$.

\subsection{Outline}

The structure of the paper is as follows. Section~\ref{sec:gs} expands
upon the graded symmetry approach to include the couplings of
flavor-singlet baryons to octet mesons and baryons, and the Lagrangian
is established.  In Section~\ref{sec:bl}, particle labels for baryons
containing a single ghost quark are explicitly written.  This partly
follows the method described in Ref.~\cite{Chen:2001yi}, with a few
differences in the notation, which are introduced for convenience in
identifying the quantum numbers and symmetry properties of the baryons
considered.  As a relevant test case, the flavor-singlet component of
the $\Lambda(1405)$ is described in Sec.~\ref{sec:ex}.

\section{Graded symmetry formalism}
\label{sec:gs}

In this section, the details of the graded symmetry approach are
briefly reviewed.  In order to derive couplings to flavor-singlet
baryons, the standard graded symmetry approach
requires a complementary augmentation in order to include interactions
between octet and singlet baryons, octet mesons, and their ghost
counterparts.

For 2+1 flavor dynamical-fermion lattice QCD calculations, the relevant
graded group is $SU(6|3)_V$, composed of three valence-quark flavors,
three sea-quark flavors and three ghost-quark flavors
\cite{Chen:2001yi}.  While the sea quark masses are free to differ
from the valence sector, the ghost sector masses are matched to the
valence sector such that their fermion-determinant contributions
exactly cancel the fermion-determinant contributions of the valence
sector, thus ensuring the valence sector is truly valence only.

However, a common approach used in practice, when separating connected
and disconnected contributions, is to work within the simpler graded
group $SU(3|3)_V$ of quenched QCD.  Here the standard quark sector is
complemented by a ghost quark sector with matching quark masses to
render the quark sector to a valence quark sector.  Identification of
the ghost sector contributions to quark-flow diagrams is then
sufficient to identify the sea-quark loop contribution of the standard
quark sector that is being removed.  This approach is used herein.

We note, however, that there are important differences between the two
graded groups, particularly for the flavor-singlet $\eta^\prime$
meson.  In $SU(3|3)_V$ of quenched QCD, the $\eta^\prime$ appears only
as a double-hairpin diagram, whereas in $SU(6|3)_V$ the $\eta^\prime$
appears in a partially-quenched generalization of the standard
$\eta^\prime$.  In our case of interest, it is sufficient to hold the
masses of each each flavor in the valence, sea and ghost sectors
equal, such that all mesons and baryons participating in the diagrams
take their standard properties. 

In deriving a Lagrangian that couples flavor-singlet baryons to octet
baryons and mesons under the graded symmetry subgroup $SU(3|3)_{V}$,
the constructions introduced in Ref.~\cite{Labrenz:1996jy} will be
used as much as possible, so that the extension to singlet couplings
can draw on the previously established formalism.

In the graded symmetry approach, the quark fields are extended to
include bosonic ghost counterparts for each flavor.  The new field, $Q
= (u,d,s,\tilde{u},\tilde{d},\tilde{s})^{\rm T}$, transforms under the vector
subgroup of the graded symmetry group $SU(3|3)_{V}$ as a fundamental
representation, analogous to three-flavor chiral
perturbation theory ($\chi$PT)
\begin{equation}
\label{eq:q}
Q_i \rightarrow U_{ij}\, Q_j, \qquad \overline{Q}_i\rightarrow \overline{Q}_j\, U^\dagger_{ji},
\end{equation}
where the indices $i,j$ run from 1 to 6.

\subsection{Mesons}

The meson field is expanded to a $6\times6$ matrix, 
with four blocks of mesons corresponding to the ordinary mesons, 
$\pi\, (q\overline{q})$, the entirely-ghost mesons, 
$\tilde{\pi}\,(\tilde{q}\overline{\tilde{q}})$, and the 
fermionic composite mesons $\chi\,(\tilde{q}\overline{q})$ and 
$\chi^\dagger\,(q\overline{\tilde{q}})$ 
in the off-diagonal blocks
\begin{equation}
\Phi = \begin{bmatrix}
\pi & \chi^\dagger \\
\chi & \tilde{\pi}
\end{bmatrix}.
\end{equation}
The elements of the ordinary meson block follow the standard
convention
\begin{equation}
\pi = \frac{1}{\sqrt{2}}\begin{pmatrix}
\frac{1}{\sqrt{2}}\pi^0+\frac{1}{\sqrt{6}}\eta & \pi^+
& K^+ \\
\pi^-& -\frac{1}{\sqrt{2}}\pi^0+\frac{1}{\sqrt{6}}\eta & 
K^0 \\
K^- & \overline{K}^0 & 
-\frac{2}{\sqrt{6}}\eta
\end{pmatrix} + \frac{\eta'}{\sqrt{6}}\mathbb{I}_3.
\end{equation}
With the inclusion of ghost quarks, the flavor of the ghost quark or
antiquark may specified through either the usual quark mixing, to form
$\tilde{\pi}^0$, $\tilde{\eta}$ and $\tilde{\eta}'$ particles, or
through neutral meson labels $\tilde{\eta}_{u}$, $\tilde{\eta}_{d}$
and $\tilde{\eta}_{s}$.  
Both representations have their merit.  When the full flavor symmetry
of the meson dressing is present, the standard representation with
$\tilde{\pi}^0$, $\tilde{\eta}$ and $\tilde{\eta}'$ is of particular
utility as the mass of the intermediate meson is manifest.  Therefore,
we consider the following form for $\chi^\dagger$
\begin{equation}
\chi^\dagger = \frac{1}{\sqrt{2}}\begin{pmatrix}
\frac{1}{\sqrt{2}}\tilde{\pi}^0+\frac{1}{\sqrt{6}}\tilde{\eta} & \tilde{\pi}^+
& \tilde{K}^+ \\
\tilde{\pi}^- & -\frac{1}{\sqrt{2}}\tilde{\pi}^0+\frac{1}{\sqrt{6}}\tilde{\eta} & 
\tilde{K}^0 \\
\tilde{K}^- & \overline{\tilde{K}}^0 & 
-\frac{2}{\sqrt{6}}\tilde{\eta}
\end{pmatrix} + \frac{\tilde{\eta}'}{\sqrt{6}}\mathbb{I}_3 \, .
\label{eq:chi}
\end{equation}
On the other hand, the flavor symmetry can be incomplete. In this
case, it is more transparent to work with the neutral $\tilde{\eta}_u$,
$\tilde{\eta}_d$ and $\tilde{\eta}_s$ mesons along the diagonal of
Eq.~(\ref{eq:chi}).  This is useful when considering
the ghost sector of a flavor-octet baryon transition to a flavor-singlet
baryon and an octet meson.  

The group transformation properties of the meson field can be made
compatible with the baryon fields via the methods of $\chi$PT, where
the exponential of the mesons is incorporated into an axial-vector
field
\begin{align}
\xi(x) &= \mathrm{exp}(i\Phi(x)/f_\pi),\\
A_\mu &= \frac{i}{2}(\xi\partial_\mu\xi^\dagger-\xi^\dagger\partial_\mu\xi).
\end{align}
The elements of $A^\mu$ transform under the graded-symmetry subgroup $SU(3|3)_{V}$ as
\begin{equation}
\label{eq:Atrans}
A^\mu_{ij}\rightarrow U_{ii'}\, A_{i'j'}^\mu\, U^\dagger_{j'j}.
\end{equation}

\subsection{Octet and singlet baryons}

In the graded symmetry approach, the baryon field is generalized to a
rank-3 tensor, allowing the three constituent flavors of each baryon
to be specified separately by each flavor index.  This is convenient,
as it readily allows one to identify the valence quark content of a
baryon simply by inspection.
The introduction of graded symmetry into the transformation matrices,
$U$, in Eq.~(\ref{eq:q}) requires the introduction of a grading factor,
defined as
\begin{equation}
\eta_i = \begin{cases} 1\quad\mbox{if}\,\,\,i=1\mbox{ -- }3\\
  0\quad\mbox{if}\,\,\,i=4\mbox{ -- }6
\end{cases},
\end{equation}
which takes into account the bosonic nature of the ghost quarks -- 
swapping two quark fields incurs a minus sign only if both are 
anticommuting (Dirac-spinor field) quarks.

The transformation properties of the octet baryon field can be  
derived from the standard nucleon interpolating field
\begin{equation}
\mathcal{B}_{ijk}^\gamma \sim [Q_i^{\alpha,a} Q_j^{\beta,b} Q_k^{\gamma,c} - 
Q_i^{\alpha,a} Q_j^{\gamma,c} Q_k^{\beta,b}]\epsilon_{abc}
(C\,\gamma_5)_{\alpha\beta} \, , 
\label{eq:octetField}
\end{equation}
where $a$ -- $c$ are color indices, $\alpha$ -- $\gamma$ are 
Dirac indices, and $C$ is the standard charge conjugation matrix. 
This construction ensures the antisymmetry of the latter two 
Dirac indices ($\beta,\gamma$), and thus eliminates the flavor-singlet
baryon \cite{Labrenz:1996jy}.

The pure flavor-singlet baryon field, denoted $\mathcal{B}^S$, on the
other hand, is totally symmetric in quark flavor, $Q$.  This can be
encoded into the form of the baryon tensor by permuting the color and
Dirac indices while maintaining the flavor-label ordering ($ijk$) as
in Eq.~(\ref{eq:octetField}), such that the established baryon transformation
rules are maintained
\begin{align}
\mathcal{B}^{S\,\,\gamma}_{ijk}\! &\sim \!\! \left [ 
  Q_i^{\alpha,a}\, Q_j^{\beta,b }\, Q_k^{\gamma,c} 
+ Q_i^{\beta,b }\, Q_j^{\gamma,c}\, Q_k^{\alpha,a} 
+ Q_i^{\gamma,c}\, Q_j^{\alpha,a}\, Q_k^{\beta,b} 
  \right . \nonumber\\
&+ \left . \,
   Q_i^{\beta,b }\, Q_j^{\alpha,a}\, Q_k^{\gamma,c} 
 + Q_i^{\gamma,c}\, Q_j^{\beta,b }\, Q_k^{\alpha,a} 
 + Q_i^{\alpha,a}\, Q_j^{\gamma,c}\, Q_k^{\beta,b} 
\right ] \nonumber\\
&\times
\epsilon_{abc}\, (C\,\gamma_5)_{\alpha\beta}.
\label{eq:singletField}
\end{align}

By transforming the individual quark operators ($Q$) within the baryon
fields of Eqs.~(\ref{eq:octetField}) and (\ref{eq:singletField}), and
collecting the transformation matrices ($U$) together using the
commutation rule \cite{Beane:2002vq}
\begin{equation}
Q_i U_{jk} = (-1)^{\eta_{i} (\eta_{j}+\eta_{k})}\, U_{jk}\, Q_i\, , 
\end{equation}
the resultant transformation rules for the baryon and antibaryons 
(suppressing the Dirac indices) are
\begin{align}
\label{eq:Btrans}
\mathcal{B}_{ijk} &\rightarrow (-1)^{\eta_{i'}(\eta_{j}+\eta_{j'})+(\eta_{i'}+\eta_{j'})(\eta_{k}+\eta_{k'})}
U_{ii'} U_{jj'} U_{kk'} \mathcal{B}_{i'j'k'}\, , \\
\overline{\mathcal{B}}_{kji} &\rightarrow (-1)^{\eta_{i'}(\eta_{j}+\eta_{j'})+(\eta_{i'}+\eta_{j'})(\eta_{k}+\eta_{k'})}
\overline{\mathcal{B}}_{k'j'i'} 
U^\dagger_{k'k} 
U^\dagger_{j'j} 
U^\dagger_{i'i} 
\, ,
\label{eq:Bbartrans}
\end{align}
for both $\mathcal{B}$ and $\mathcal{B}^S$.

In matching the elements of the generalized baryon field to those of
the standard 
baryon matrix $B_{kk'}$, the indices are restricted from $1$ to $3$,
and the elements are assigned so that the result is antisymmetric
under $i \leftrightarrow j$
\begin{equation}
\label{eq:bary}
\mathcal{B}_{ijk} \propto \epsilon_{ijk'}B_{kk'}, \quad (i\mbox{ -- }k = 
1\mbox{ -- }3), 
\end{equation}
where the convention used for the standard baryon field of $\chi$PT is
\begin{equation}
\label{eq:chiPTbaryons}
B = \begin{pmatrix}
\frac{1}{\sqrt{2}}\Sigma^0 + \frac{1}{\sqrt{6}}\Lambda & \Sigma^{+} & p \\
\Sigma^{-} & -\frac{1}{\sqrt{2}}\Sigma^0 + \frac{1}{\sqrt{6}}\Lambda & n \\
\Xi^{-} & \Xi^{0} & -\frac{2}{\sqrt{6}}\Lambda 
\end{pmatrix} + \frac{\Lambda'}{2\sqrt{3}} \mathbb{I}_3.
\end{equation}
$\Lambda'$ denotes a positive-parity flavor-singlet baryon.
Similarly, a matrix $B^*$ can be defined with the same form,
representing the lowest-lying negative-parity baryons.  The
negative-parity flavor-singlet baryon is then denoted by
$\Lambda^{\prime *}$.

The baryon construction of Eq.~(\ref{eq:bary}) contains both octet
and singlet fields. However, it is known that interactions that
include a singlet baryon have couplings that are not related via
flavor symmetry to the interaction couplings between octet baryons
only.  It is therefore important to separate the baryon field into
`octet-only' and `singlet-only' pieces, so that the terms in the
Lagrangian can be assigned independent coefficients.

To extract the purely octet component of $\mathcal{B}_{ijk}$, Labrenz
and Sharpe antisymmetrized over the $j$ and $k$ indices of $Q$ in
Eq.~(\ref{eq:octetField}), so that the symmetric component (containing
the singlet) is removed.  In terms of the standard 
baryon field, with indices $i,\ j,$ and $k$ restricted to $1$ -- $3$ and
fermion anticommutation taken into account, this takes the following
form
\begin{equation}
\label{eq:B}
\mathcal{B}_{ijk} = \frac{1}{\sqrt{6}}\left(
\epsilon_{ijk'}B_{kk'} + \epsilon_{ikk'}B_{jk'}\right). 
\end{equation}
The factor of $1/\sqrt{6}$ is chosen so that the baryon wave functions
have the correct normalization. This becomes apparent when assigning
conventional baryon labels to the elements of the $\mathcal{B}$ tensor
(discussed in Sec~\ref{sec:bl}). For example, extracting the proton
states involves selecting the elements $\mathcal{B}_{112}$,
$\mathcal{B}_{121}$ and $\mathcal{B}_{211}$, with weights of
$1/\sqrt{6}$, $1/\sqrt{6}$ and $-2/\sqrt{6}$, respectively, whose
squares sum to 1.

Similarly, the baryon tensor containing the flavor-singlet baryon
follows from Eq.~(\ref{eq:singletField}) for $\mathcal{B}^S$
\begin{align}
\label{eq:BS}
\mathcal{B}^S_{ijk} = 
\frac{1}{3\sqrt{2}} \big ( &
 +\epsilon_{ijk'}B_{kk'} 
 +\epsilon_{jkk'}B_{ik'}
 +\epsilon_{kik'}B_{jk'}
\nonumber\\
&-\epsilon_{jik'}B_{kk'} 
 -\epsilon_{kjk'}B_{ik'}
 -\epsilon_{ikk'}B_{jk'} 
\big ) \, . 
\end{align}
Again, the normalization factor of $(3\sqrt{2})^{-1}$ ensures the squares of
the coefficients sum to 1.  
For the $\Lambda'$, one considers the six
terms, $
\mathcal{B}^S_{123},\ 
\mathcal{B}^S_{132},\ 
\mathcal{B}^S_{213},\ 
\mathcal{B}^S_{231},\ 
\mathcal{B}^S_{312}\ {\rm and}\ 
\mathcal{B}^S_{321}
$, each of which provides a contribution with magnitude $\Lambda'/\sqrt{6}$. 
Similarly
\begin{equation}
\label{eq:Lprime}
\epsilon_{ijk}\, \mathcal{B}^S_{ijk} = \sqrt{6}\,\Lambda',\quad
(i\mbox{ -- }k = 1\mbox{ -- }3) \, .
\end{equation}

The symmetry relations that describe the relationships among the
elements of the baryon tensors are important when considering their
general form in the graded symmetry group, {\it i.e.} where the indices
$i$ -- $k$ run from $1$ to $6$.  For the octet field $\mathcal{B}$, the
following two symmetry relations are sufficient to specify the two
group invariants of the symmetry, $\alpha$ and $\beta$
\cite{Labrenz:1996jy}, which are related to the standard $D$ and $F$
couplings
\begin{subequations}
\begin{align}
\label{eq:Blattertwo}
\mathcal{B}_{ikj} &= (-1)^{\eta_j\eta_k+1}\mathcal{B}_{ijk}\, , \\
\mathcal{B}_{ijk} &= (-1)^{\eta_i\eta_j}\mathcal{B}_{jik}
+(-1)^{\eta_i\eta_j+\eta_j\eta_k+\eta_k\eta_i}\mathcal{B}_{kji} \, .
\label{eq:Bsym2}
\end{align}
\end{subequations}
Similarly, the following symmetry relations exist for the field
$\mathcal{B}^S$
\begin{subequations}
\begin{align}
\mathcal{B}^S_{jik} =& 
(-1)^{\eta_i\eta_j}\, \mathcal{B}^S_{ijk}  \, , \label{eq:BSsym1}\\
\mathcal{B}^S_{ikj} =& 
(-1)^{\eta_j\eta_k}\, \mathcal{B}^S_{ijk} \, , \label{eq:BSlattertwo}\\
\mathcal{B}^S_{jki} =& 
(-1)^{\eta_i\eta_k + \eta_i\eta_j}\, \mathcal{B}^S_{ijk} \, , \label{eq:BSsym3}\\
\mathcal{B}^S_{kij} =& 
(-1)^{\eta_i\eta_k + \eta_j\eta_k}\, \mathcal{B}^S_{ijk} \, , \label{eq:BSsym4}\\
\mathcal{B}^S_{kji} =& 
(-1)^{\eta_i\eta_j + \eta_i\eta_k + \eta_j\eta_k}\,
\mathcal{B}^S_{ijk} \, . \label{eq:BSsym5}
\end{align}
\end{subequations}

\subsection{The Lagrangian}
\label{subsec:lag}

The singlet Lagrangian will couple the singlet-baryon field to a
combination of octet-meson and octet-baryon fields, 
$A\cdot\mathcal{B}$.
Given the maximal symmetry for ${\mathcal{B}}^S$ in
Eq.~(\ref{eq:singletField}), there are three different ways of
contracting the octet-baryon flavor indices.  In keeping with the
baryon construction of Labrenz and Sharpe \cite{Labrenz:1996jy}, we
consider a Lagrangian term similar to their term with coefficient
$\alpha$,
$\overline{\mathcal{B}}^S_{kji}\, A_{kk'}\, \mathcal{B}_{ijk'}$,
where the first two indices of $\mathcal{B}$ are contracted
with $\ca{B}^S$;
a term similar to their term with coefficient  $\beta$,
$\overline{\mathcal{B}}^S_{kji}\, A_{ik'}\, \mathcal{B}_{k'jk}$, 
where the latter two indices of $\mathcal{B}$ are contracted
with $\ca{B}^S$;
and a third candidate 
$\overline{\mathcal{B}}^S_{kji}\, A_{kk'}\, \mathcal{B}_{ik'j}$,
where the outer indices of $\mathcal{B}$ are contracted with
$\ca{B}^S$.  Noting that the octet field of Eq.~(\ref{eq:octetField})
has a symmetry constraint under the exchange of the latter
two indices as in Eq.~(\ref{eq:Blattertwo}), this third candidate is
related to the first candidate and may be discarded.  
Similarly, the relation of {Eq.~(\ref{eq:BSsym1}) for $\ca{B}^S$ is
opposite to that of Eq.~(\ref{eq:Blattertwo}) for $\mathcal{B}$ and
therefore the second candidate vanishes.
Thus, there is only one group invariant combination.

In the general case where all indices run from $1$ -- $6$, the
commutation/anticommutation behavior of the elements of the
transformation matrices $U$ depends on their location in the block
matrix.  As a result, grading factors must be included when swapping
the order of the indices.
Noting that Eq.~(\ref{eq:BSsym5}) indicates that $\mathcal{B}^S_{ijk}$ is antisymmetric under 
exchange of the outer indices in both the normal quark and single ghost-quark sectors,
the interaction Lagrangians take the following forms
\begin{align}
\label{eq:LagS}
\mathcal{L}_S &= 2 \sqrt{2}\, g_s
\,(-1)^{(\eta_i+\eta_j)(\eta_k+\eta_{k'})}\, \overline{\mathcal{B}}^S_{kji}\,
A_{kk'}\, \mathcal{B}_{ijk'}, \\
\mathcal{L}_{SS} &= \sqrt{6}\, g_{ss} 
\,(-1)^{(\eta_i+\eta_j)(\eta_k+\eta_{k'})}\, \overline{\mathcal{B}}^S_{kji}\,
A_{kk'}\, \mathcal{B}^S_{ijk'}. 
\label{eq:LagSS}
\end{align}
We note these are special cases of the general irreducible
representations of graded Lie groups classified by Balantekin and Bars
\cite{BahaBalantekin:1980tup,BahaBalantekin:1980igq}.
The leading coefficient in $\mathcal{L}_S$ is selected to define the
coupling associated with the process $\Lambda'\rightarrow
\pi_0\Sigma_0$ as $g_s$.

Our investigation of the separation of quark-flow connected and
disconnected contributions to standard baryons constrains the indices
$i$ -- $k$ to $1$ -- $3$, whereas $k'=1$ -- $6$ incorporates the ghost
meson-baryon contributions, which reveal the disconnected
contributions.

In summary, the first-order interaction Lagrangian for low-lying octet
and singlet mesons, octet baryons and low-lying negative-parity
singlet baryons, including the Dirac structure, takes the form
\begin{align}
\mathcal{L}^{(1)}_{\mathcal{B}A} =& 
\, \mathcal{L}_\alpha + \mathcal{L}_\beta + \left ( \mathcal{L}_S +
\mathrm{h.c.} \right ) + \mathcal{L}_{SS} \, , \nonumber \\
=&\, 2\alpha\, (-1)^{(\eta_i+\eta_j)(\eta_k+\eta_{k'})}\,
\overline{\mathcal{B}}_{kji}\, \gamma_\mu \gamma_5 \,
A^\mu_{kk'}\, \mathcal{B}_{ijk'} \nonumber\\
+&\, 2\beta\, 
\overline{\mathcal{B}}_{kji}\, \gamma_\mu \gamma_5 \,
A^\mu_{ik'}\, \mathcal{B}_{k'jk} \nonumber\\
+&\, 2 \sqrt{2}\, g_s \, (-1)^{(\eta_i+\eta_j)(\eta_k+\eta_{k'})}\, \nonumber \\
&\times
\left \{ \overline{\mathcal{B}}^{S*}_{kji}\, \gamma_\mu\, A^\mu_{kk'}\,
\mathcal{B}_{ijk'} 
+ \overline{\mathcal{B}}_{k'ji}\, \gamma_\mu\, A^\mu_{k'k}\,
\mathcal{B}^{S*}_{ijk} \right \} \nonumber\\
+&\sqrt{6}\, g_{ss} \,(-1)^{(\eta_i+\eta_j)(\eta_k+\eta_{k'})}\, \overline{\mathcal{B}}^{S*}_{kji}\,
\gamma_\mu \gamma_5 \,A^\mu_{kk'}\, \mathcal{B}^{S*}_{ijk'} \, ,
\label{eq:myl}
\end{align}
where $\mathcal{B}^{S*}_{ijk}$ denotes the odd-parity generalized
field associated with $B^*$ and thus $\Lambda^{\prime *}$ via
Eq.~(\ref{eq:BS}).  The loss of $\gamma_5$ in the terms coupling
octet- and singlet-flavor baryons reflects the odd parity of the
low-lying $\Lambda^{\prime *}$.

\section{Ghost Baryon State Identification}
\label{sec:bl}

In this section we review briefly the process of relating the elements
of $\mathcal{B}$ and $\mathcal{B}^S$ to baryons having one ghost quark
and two ordinary quarks.  This is relevant to our investigation of the
ordinary $\Lambda^{\prime *}$ baryon, where one ghost quark can be introduced
through the disconnected sea-quark loop.  The ordinary baryons (those
without ghost quark content) adopt the labeling convention assigned in
Eqs.~(\ref{eq:chiPTbaryons}) through (\ref{eq:BS}).  The case for one
ghost quark in $\mathcal{B}$ was first present by Chen and Savage
\cite{Chen:2001yi}.  We will extend the formalism to include
$\mathcal{B}^S$. 

In identifying the intermediate states of the meson-baryon channels
dressing a baryon, we find it is useful to assign state labels that
provide information about the charge, quark composition and flavor
symmetry.
Thus, in defining the baryon labels, there are some differences in our
notation with respect to Ref.~\cite{Chen:2001yi}. Here, the choice in
baryon labels is to manifest the symmetry and flavor properties of
baryons with a single ghost quark.  If one of the quarks is a
(commuting) ghost particle, it may be combined with the two remaining
ordinary quarks in a $\Sigma$-like symmetric or a $\Lambda$-like
antisymmetric configuration of ordinary quarks.  For example, for a
proton with a ghost up quark, one defines new particles
$\tilde{\Sigma}^{+}_{p,\tilde{u}}$, and
$\tilde{\Lambda}^+_{p,\tilde{u}}$, which correspond to symmetric and
antisymmetric combinations of the two remaining quark flavors,
respectively.  

In addition, unphysical states such as doubly-charged `protons'
$\tilde{\Sigma}^{++}_{p,\tilde{u}}\,(\tilde{u},u,u)$ and negatively
charged `neutrons' $\tilde{\Sigma}^-_{n,\tilde{d}}\,(\tilde{d},d,d)$
are members of the single ghost-quark baryons.  These states occur in
both the graded-symmetry approach and the diagrammatic approach
by considering all possible quark-flow topologies.  Note that these
states do not contribute in full QCD, where the cancellation of the
connected and disconnected quark-flow diagrams associated with them
enforces the Pauli exclusion principle.

Following Ref.~\cite{Chen:2001yi}, the flavor-symmetric variants are
encoded in a rank-3 tensor $s_{ijk}$, which is symmetric in the last
two indices reserved for ordinary quarks
\begin{align}
s_{411}&=\tilde{\Sigma}^{++}_{p,\tilde{u}},\hspace{4pt} &&s_{511}=\tilde{\Sigma}^{+}_{p,\tilde{d}}, 
\hspace{4pt} &&s_{611}=\tilde{\Sigma}^{+}_{\Sigma,\tilde{s}}, \nonumber\\
s_{412}&=\tilde{\Sigma}^{+}_{p,\tilde{u}}/\sqrt{2},
\hspace{4pt} &&s_{512}=\tilde{\Sigma}^{0}_{n,\tilde{d}}/\sqrt{2},
\hspace{4pt} &&s_{612}=\tilde{\Sigma}^{0}_{\Sigma,\tilde{s}}/\sqrt{2},\nonumber\\
s_{422}&=\tilde{\Sigma}^{0}_{n,\tilde{u}},\hspace{4pt} &&s_{522}=\tilde{\Sigma}^{-}_{n,\tilde{d}}, 
\hspace{4pt} &&s_{622}=\tilde{\Sigma}^{-}_{\Sigma,\tilde{s}}, \nonumber\\
s_{413}&=\tilde{\Sigma}^{+}_{\Sigma,\tilde{u}}/\sqrt{2},
\hspace{4pt} &&s_{513}=\tilde{\Sigma}^{0}_{\Sigma,\tilde{d}}/\sqrt{2},
\hspace{4pt} &&s_{613}=\tilde{\Sigma}^{0}_{\Xi,\tilde{s}}/\sqrt{2},\nonumber\\
s_{423}&=\tilde{\Sigma}^{0}_{\Sigma,\tilde{u}}/\sqrt{2},
\hspace{4pt} &&s_{523}=\tilde{\Sigma}^{-}_{\Sigma,\tilde{d}}/\sqrt{2},
\hspace{4pt} &&s_{623}=\tilde{\Sigma}^{-}_{\Xi,\tilde{s}}/\sqrt{2},\nonumber\\
s_{433}&=\tilde{\Sigma}^{0}_{\Xi,\tilde{u}},\hspace{4pt} &&s_{533}=\tilde{\Sigma}^{-}_{\Xi,\tilde{d}}, 
\hspace{4pt} &&s_{633}=\tilde{\Sigma}^{-}_{\Omega,\tilde{s}}. 
\end{align}
The baryon subscripts reflect the combined number of strange quarks and
strange ghost quarks and the superscripts record the total electric
charge.  The flavor of the residing ghost quark is explicit in the
subscript.  We note that although $\Omega$ has been used to denote the
$(\tilde{s},s,s)$ combination, this baryon is actually associated with
octet as opposed to decuplet baryons.

Using a similar notation, the $\Lambda$-like particles are encoded in
an antisymmetric tensor, denoted $t_{ij}$
\begin{align}
t_{41} &= \tilde{\Lambda}^0_{\Sigma,\tilde{u}},
\quad &&t_{51} = \tilde{\Lambda}^-_{\Sigma,\tilde{d}},
\quad &&t_{61} = \tilde{\Lambda}^-_{\Xi,\tilde{s}},\nonumber\\
t_{42} &= \tilde{\Lambda}^+_{\Sigma,\tilde{u}},
\quad &&t_{52} = \tilde{\Lambda}^0_{\Sigma,\tilde{d}},
\quad &&t_{62} = \tilde{\Lambda}^0_{\Xi,\tilde{s}},\nonumber\\
t_{43} &= \tilde{\Lambda}^+_{p,\tilde{u}},
\quad &&t_{53} = \tilde{\Lambda}^0_{n,\tilde{d}},
\quad &&t_{63} = \tilde{\Lambda}^0_{\Sigma,\tilde{s}}. 
\label{eq:antisymmbar}
\end{align}

These baryons are distributed over the elements of $\mathcal{B}$ in
the following manner.  The manifestly symmetric part of $\mathcal{B}$
has the simple relation
\begin{equation}
\mathcal{B}_{ijk} = \sqrt{\frac{2}{3}}s_{ijk}, \quad (i=4\mbox{ --
}6,\,\, j\mbox{ -- }k = 
1\mbox{ -- }3),
\label{eq:symmetric mapping}
\end{equation}
which places the ghost quark in the first index position of
$\mathcal{B}$.  The symmetric states with ghost quarks mapped to the
second and third index positions of $\mathcal{B}$ are combined with
the antisymmetric baryons of Eq.~(\ref{eq:antisymmbar}) in
\begin{subequations}
\label{eq:antisymmetric mapping}
\begin{align}
\mathcal{B}_{jik} &= \frac{1}{2}\, t_{ip}\, \epsilon_{pjk} + \frac{1}{\sqrt{6}}\, s_{ijk} \, , \\
\mathcal{B}_{jki} &= -\mathcal{B}_{jik} \, , 
\end{align}
\end{subequations}
where $i=4$ -- $6$, $j$ -- $k=1$ -- $3$ and $p=1$ -- $3$.  
These two equations map the
symmetric baryon states to two elements of $\mathcal{B}$ with weight
$1/\sqrt{6}$ such that the weighting of $\sqrt{2/3}$ in
Eq.~(\ref{eq:symmetric mapping}) provides a sum of squares equal to
one.  Similarly, Eqs.~(\ref{eq:antisymmetric mapping}) distribute the
antisymmetric ghost baryon states of Eqs.~(\ref{eq:antisymmbar}) over
four elements of $\mathcal{B}$ with weight $1/2$ such that a sum of
squares again equals one.

The complete flavor antisymmetry of $\mathcal{B}^S$ for ordinary quarks
restricts its first ghost sector to baryon states having antisymmetry
in the ordinary quarks, {\it i.e.} the baryons of
Eq.~(\ref{eq:antisymmbar})
\begin{subequations}
\label{eq:BSmapping}
\begin{align}
\mathcal{B}^S_{ijk} &= \frac{1}{\sqrt{6}}\, t'_{ip}\, \epsilon_{pjk} \, ,\\
\mathcal{B}^S_{jik} &= +\mathcal{B}^S_{ijk} \, , \\
\mathcal{B}^S_{jki} &= +\mathcal{B}^S_{ijk} \, .
\end{align}
\end{subequations}
where $i=4$ -- $6$ and $j$ -- $k=1$ -- $3$.  
Eqs.~(\ref{eq:BSmapping}), now with a prime to denote the 
association with the flavor-singlet baryon sector, maps 
each baryon of Eqs.~(\ref{eq:antisymmbar}) over six elements of
$\mathcal{B}^S$.  
Analogous results are obtained for the odd-parity
$\mathcal{B}^{S*}_{ijk}$ using $t^{\prime *}_{ip}$.

In studies of the $\Lambda^{\prime *}$, the low-lying ghost states
reside in the octet meson and octet baryon sectors.
The odd-parity ghost states of $\mathcal{B}^{S*}_{ijk}$ are of
relevance to other states transiting to $\Lambda^{\prime *}$ and a
meson.  As an example, consider an octet baryon transition to a
singlet baryon and an octet meson.  Such physics is relevant to
understanding the disconnected loop contributions in $\Sigma
\rightarrow \pi \, \Lambda^{\prime *}$ or $\Xi \rightarrow \overline
K\, \Lambda^{\prime *}$.  These become particularly important in
lattice QCD, where the initial state can be a finite-volume excitation
associated with a resonance.

This defines the method by which baryon labels are assigned to
different elements of the three-index baryon tensors, $\ca{B}$ and
$\ca{B}^S$, and their odd-parity analogues.  
By writing out the full Lagrangian in Eq.~(\ref{eq:myl}),
one can simply pick out the relevant contributions to each baryon in
each interaction channel available for the symmetry 
subgroup $SU(3|3)_{V}$.

\section{Flavor-singlet $\mathbf{\Lambda}$-baryon couplings}\label{sec:ex}

In this section we examine the coupling strengths for transitions
involving the flavor-singlet component, $\Lambda^{\prime *}$.  As
described at the beginning of Sec.~\ref{sec:gs}, the couplings of the
ghost sector in $SU(3|3)_{V}$ provide the couplings of quark-flow
disconnected sea-quark loop contributions in $SU(6|3)_{V}$ that we are
seeking.  In writing out the relevant terms of the Lagrangian, we
report the meson-baryon $\rightarrow$ baryon transition terms, noting
that the baryon $\rightarrow$ meson-baryon terms are given by the
Hermitian conjugate.

\subsection{Flavor-octet baryon dressings of the singlet baryon}\label{subsec:s2o}

Here we consider the specific case relevant to the low-lying
odd-parity $\Lambda(1405)$ baryon, known to have an important
flavor-singlet component \cite{Menadue:2011pd}.  We commence by
considering the process $\Lambda^{\prime *} \rightarrow
\mathcal{B}\,\Phi \rightarrow \Lambda^{\prime *}$ via the Lagrangian
of Eq.~(\ref{eq:myl}).

In this case, the $u$, $d$ and $s$ valence quarks that comprise the
$\Lambda^{\prime *}$ state may each contribute to the $\Phi$ meson
loop, shown in Fig.~1.  The ghost quark sector may enter through the
disconnected loop, and therefore there are three choices for the
antiquark plus three choices for the ghost antiquark.  Noting that the
flavor-singlet $\eta'$ meson cannot be combined with an octet baryon
to form a flavor singlet, there are $(3\times6) - 1 = 17$ terms in the
Lagrangian for the $\overline{\Lambda}^{\prime *}\, \Phi\, \mathcal{B}$ coupling.
At one loop, the baryon $\mathcal{B}$ contains at most one ghost quark
for which the meson contains the corresponding antighost quark.

The $SU(3|3)_{V}$-flavor composition of the interaction Lagrangian of
Eq.~(\ref{eq:myl}) contains the regular full-QCD meson-baryon
couplings to the singlet $\Lambda^{\prime *}$
\begin{align}
\mathcal{L}_{S,\overline{\Lambda}^{\prime *}}^{\rm QCD} = 
g_s\, \overline{\Lambda}^{\prime *} \, \Big\{\, &
   \overline{K}^0\, n 
 + K^-\, p \nonumber \\
+&\, \pi^+\, \Sigma^- 
    +\pi^0\, \Sigma^0 
    +\pi^-\, \Sigma^+ \nonumber \\
+&\,  K^+\, \Xi^- 
    + K^0\, \Xi^0 
    + \eta\, \Lambda \,
 \Big \} \, ,
\label{eq:fullQCD}
\end{align}
and the ghost meson to ghost baryon couplings
\begin{align}
\mathcal{L}_{S,\overline{\Lambda}^{\prime *}}^{\rm Ghost} = 
-g_s\, \overline{\Lambda}^{\prime *}\, \bigg \{ \, &
\sqrt{\frac{2}{3}}\, \Big ( \ 
    \tilde{K}^-\,\tilde{\Lambda}^+_{p,\tilde{u}} 
  + \overline{\tilde{K}}^0\,\tilde{\Lambda}^0_{n,\tilde{d}}  \nonumber \\
&\qquad
  + \tilde{\pi}^+\, \tilde{\Lambda}^-_{\Sigma,\tilde{d}} \,
  + \tilde{\pi}^-\, \tilde{\Lambda}^+_{\Sigma,\tilde{u}}  \nonumber \\
&\qquad
  + \tilde{K}^+\,\tilde{\Lambda}^-_{\Xi,\tilde{s}} 
  + \tilde{K}^0\,\tilde{\Lambda}^0_{\Xi,\tilde{s}} 
\Big ) \nonumber \\
& + \frac{1}{\sqrt{3}}\Big( \,
    \tilde{\pi}_0 \, \tilde{\Lambda}^0_{\Sigma,\tilde{u}}
  - \tilde{\pi}_0 \, \tilde{\Lambda}^0_{\Sigma,\tilde{d}} 
\, \Big ) \nonumber\\
& + \frac{1}{3} \, \Big ( \ 
       \tilde{\eta}\,\tilde{\Lambda}^0_{\Sigma,\tilde{u}} 
  +    \tilde{\eta}\,\tilde{\Lambda}^0_{\Sigma,\tilde{d}}  
  -2\, \tilde{\eta}\,\tilde{\Lambda}^0_{\Sigma,\tilde{s}} 
\Big )
 \bigg \},
\label{eq:PQQCD}
\end{align}
where the Dirac structure has been suppressed. 
In the full-QCD case, the relative coupling strengths of the
$\pi\Sigma$, $\overline{K}N$, $K\Xi$ and $\eta\Lambda$ channels are
consistent with early work \cite{Veit:1984an} as expected.

The ghost terms, however, present novel features.  
Only antisymmetric $\tilde\Lambda$ ghost baryons contribute.
Thus, the coupling of a baryon with a single ghost quark to a
flavor-singlet baryon is constrained to have its two ordinary quarks
in an antisymmetric formation.  The $\tilde{\Sigma}$ baryons do
not contribute.

In addition, flavor-singlet symmetry is manifest.  Each meson charge
state participating in the quark-flow disconnected loop contributes
in a uniform manner with a coupling strength of $\frac{2}{3}\, g_s^2$.

The ghost-quark subscript on the baryon field enables the
identification of the flavor of the quark participating in the loop.
Thus, matrix elements can be separated into their quark-flow connected
and disconnected parts for each quark flavor.  Similarly, the baryon
subscripts facilitate the incorporation of $SU(3)$-flavor breaking
effects through variation of the hadron masses.

The essential feature to be drawn from this analysis is that the
relative contribution of quark-flow connected and disconnected
diagrams is now known.
Consider the process
${\Lambda}^{\prime *} \rightarrow K^-\, p  \rightarrow {\Lambda}^{\prime *}$
proportional to $g_s^2$.  This process proceeds through a combination of
quark-flow connected diagrams and the quark-flow disconnected diagram
of Fig.~\ref{fig:Bloop}.  See for example, Ref.~\cite{Hall:2013dva}.
The ghost Lagrangian reports a term
\begin{equation}
- g_s\, \sqrt{\frac{2}{3}}\; \overline{\Lambda}^{\prime *}\, \tilde{K}^-\,
\tilde{\Lambda}^+_{p,\tilde{u}} \, ,
\end{equation}
indicating that this process proceeds through the quark-flow
disconnected diagram with strength $\frac{2}{3}\, g_s^2$.  That is, two
thirds of the full-QCD process
${\Lambda}^{\prime *} \rightarrow K^-\, p  \rightarrow {\Lambda}^{\prime *}$
proceeds by the quark-flow disconnected diagram with a $u$ quark
flavor participating in the loop.  The fact that two-thirds of the
strength lies in the loop is perhaps surprising.

The full-QCD processes involving the neutral $\pi^0$ or $\eta$ mesons
are only slightly more complicated, as one needs to sum over more than
one quark flavor participating in the sea-quark loop.  After summing
over the participating flavors, each full-QCD process proceeds with
one third of the strength in quark-flow connected diagrams and
two-thirds of the strength in quark-flow disconnected diagrams.

\subsection{Flavor-singlet baryon dressings of octet baryons}\label{subsec:o2s}

It is also of interest to examine processes where the flavor-singlet
baryon participates as an intermediate state dressing of an octet
baryon.  The full Lagrangian describing the processes
$\mathcal{B}\rightarrow \Phi\, \mathcal{B}^{S*} \rightarrow
\mathcal{B}$, where $\mathcal{B}$ is an octet baryon, can likewise be
decomposed into full-QCD and ghost meson to ghost baryon components,
suppressing the Dirac structure
\begin{align}
\mathcal{L}_{S,\Lambda^{\prime *}}^{\rm QCD} = 
g_s\, \Big\{\, &
   \overline{n}\, ( K^0\, \Lambda^{\prime *})
 + \overline{p}\, ( K^+\, \Lambda^{\prime *}) \nonumber \\
+&\, \overline{\Sigma}^-\, ( \pi^-\, \Lambda^{\prime *}) 
    +\overline{\Sigma}^0\, ( \pi^0\, \Lambda^{\prime *})
    +\overline{\Sigma}^+\, ( \pi^+\, \Lambda^{\prime *}) \nonumber \\
+&\,  \overline{\Xi}^-\, ( K^-\, \Lambda^{\prime *}) 
    + \overline{\Xi}^0\, ( \overline{K}^0\, \Lambda^{\prime *} ) \nonumber \\
+&\,  \overline{\Lambda}\, ( \eta\, \Lambda^{\prime *}) \,
 \Big \} \, ,
\label{eq:fullQCD2}
\end{align}

\begin{widetext}
\begin{align}
&\mathcal{L}_{S,\Lambda^{\prime *}}^{\rm Ghost} = g_s \,\bigg\{\quad
%
%
\overline{n}\,\bigg(
  \tilde{\pi}^-\,\tilde{\Lambda}^{+\prime *}_{p,\tilde{u}}
+ \tilde{\eta}_d\,\tilde{\Lambda}^{0\prime *}_{n,\tilde{d}}
+ \tilde{K}^0\,  \tilde{\Lambda}^{0\prime *}_{\Sigma,\tilde{s}} 
\bigg)
%
%
\quad + \;\quad
  \overline{p}\,\bigg(
 \tilde{\eta}_u\,\tilde{\Lambda}^{+\prime *}_{p,\tilde{u}} 
+ \tilde{\pi}^+\,\tilde{\Lambda}^{0\prime *}_{n,\tilde{d}}
+ \tilde{K}^+\,  \tilde{\Lambda}^{0\prime *}_{\Sigma,\tilde{s}} 
\bigg)\nonumber\\
%
%
&+\overline{\Sigma}^-\,\bigg(
  \tilde{\pi}^-\,\tilde{\Lambda}^{0\prime *}_{\Sigma,\tilde{u}} 
+\tilde{\eta}_d\,\tilde{\Lambda}^{-\prime *}_{\Sigma,\tilde{d}}
+ \tilde{K}^0\,  \tilde{\Lambda}^{-\prime *}_{\Xi,\tilde{s}} 
\bigg)
%
%
\quad+\quad
  \overline{\Sigma}^+\,\bigg(
  \tilde{\eta}_u\,\tilde{\Lambda}^{+\prime *}_{\Sigma,\tilde{u}} 
+ \tilde{\pi}^+\,\tilde{\Lambda}^{0\prime *}_{\Sigma,\tilde{d}}
+ \tilde{K}^+\,  \tilde{\Lambda}^{0\prime *}_{\Xi,\tilde{s}} 
\bigg)\nonumber\\
%
%
&+\frac{1}{\sqrt{2}}\,\overline{\Sigma}^0\,\bigg(
-  \tilde{\pi}^-\, \tilde{\Lambda}^{+\prime *}_{\Sigma,\tilde{u}} 
+  \tilde{\eta}_u\,\tilde{\Lambda}^{0\prime *}_{\Sigma,\tilde{u}}
                            -\tilde{\eta}_d\,\tilde{\Lambda}^{0\prime *}_{\Sigma,\tilde{d}}
 +\tilde{\pi}^+\,\tilde{\Lambda}^{-\prime *}_{\Sigma,\tilde{d}}
 -\tilde{K}^0\,  \tilde{\Lambda}^{0\prime *}_{\Xi,\tilde{s}}
 +\tilde{K}^+\,  \tilde{\Lambda}^{-\prime *}_{\Xi,\tilde{s}}
\bigg)\nonumber\\
%
%
&+\overline{\Xi}^-\,\bigg(
  \tilde{K}^-\,\tilde{\Lambda}^{0\prime *}_{\Sigma,\tilde{u}}
+ \overline{\tilde{K}}^0\,\tilde{\Lambda}^{-\prime *}_{\Sigma,\tilde{d}} 
+\tilde{\eta}_s\,\tilde{\Lambda}^{-\prime *}_{\Xi,\tilde{s}}
\bigg)
%
%
\quad + \quad
  \overline{\Xi}^0\,\bigg(
  \tilde{K}^-\,           \tilde{\Lambda}^{+\prime *}_{\Sigma,\tilde{u}}
+ \overline{\tilde{K}}^0\,\tilde{\Lambda}^{0\prime *}_{\Sigma,\tilde{d}} 
+\tilde{\eta}_s\,\tilde{\Lambda}^{0\prime *}_{\Xi,\tilde{s}}
\bigg)\nonumber\\
%
%
&+\frac{1}{\sqrt{6}}\,\overline{\Lambda}\,\bigg(
  \tilde{\pi}^-\, \tilde{\Lambda}^{+\prime *}_{\Sigma,\tilde{u}} 
+ \tilde{\eta}_u\,\tilde{\Lambda}^{0\prime *}_{\Sigma,\tilde{u}}
                            +\tilde{\eta}_d\,\tilde{\Lambda}^{0\prime *}_{\Sigma,\tilde{d}}
- 2\tilde{\eta}_s\, \tilde{\Lambda}^{0\prime *}_{\Sigma,\tilde{s}}
+ \tilde{\pi}^+\, \tilde{\Lambda}^{-\prime *}_{\Sigma,\tilde{d}}
-  2\,\tilde{K}^-\,            \tilde{\Lambda}^{+\prime *}_{p,\tilde{u}}
-  2\,\overline{\tilde{K}}^0\,\tilde{\Lambda}^{0\prime *}_{n,\tilde{d}} 
+ \tilde{K}^0\,\tilde{\Lambda}^{0\prime *}_{\Xi,\tilde{s}}
+ \tilde{K}^+\,\tilde{\Lambda}^{-\prime *}_{\Xi,\tilde{s}}
\bigg)\quad\bigg\} \, .
\label{eq:PQQCD2}
\end{align}
\end{widetext}
In each channel of Eq.~(\ref{eq:PQQCD2}), there are terms which represent the disconnected loop
contributions of the normal full-QCD interactions of Eq.~(\ref{eq:fullQCD2}), and there are
additional terms that serve to cancel the unphysical quark-flow connected diagrams associated with
each channel. For example, in the case of the neutron channel $\overline{n}$, the relevant ghost
term $\tilde{K}^0\, \tilde{\Lambda}^{0\prime *}_{\Sigma,\tilde{s}}$ comes with the same magnitude
as the normal process, $g_s$, indicating that there is no completely-connected contribution for
this process.  The physical process proceeds through a quark-flow disconnected diagram.  The other
two terms in this channel, $\tilde{\pi}^-\,\tilde{\Lambda}^{+\prime *}_{p,\tilde{u}}$ and
$\tilde{\eta}_d\,\tilde{\Lambda}^{0\prime *}_{n,\tilde{d}}$ occur to cancel the aforementioned
unphysical connected diagrams.  These contributions are also important and must be included when
evaluating the quark-flow disconnected contributions to baryon matrix elements.

Note that, unlike the flavor-singlet $\Lambda$-baryon couplings of Eq.~(\ref{eq:PQQCD}), where the
structure of the $\tilde{\pi}^0$ and $\tilde{\eta}$ is manifest, here there is no $s\overline{s}$
contribution in any channel apart from the octet $\overline{\Lambda}$.  In the $\overline{\Sigma}$, 
where both light flavor $\eta_q$ mesons appear, the flavor symmetry of the pion is apparent.
Similarly, in the $\overline{\Lambda}$ channel, the flavor symmetry of the $\eta_{q}$ contributions
identifies the octet-$\eta$ participating to cancel the full-QCD process $\overline{\Lambda}\, (
\eta\, \Lambda^{\prime *})$.  In all other cases, the light-quark $\eta_u$ and $\eta_d$ mesons
propagate with the mass of the pion.  
\color{black}

\subsection{Flavor-singlet baryon dressings of the singlet baryon}\label{subsec:s2s}

Finally, we also consider the quark-flow disconnected loop contributions to the process
$\Lambda^{\prime *} \rightarrow \Phi\, \mathcal{B}^{S*} \rightarrow \Lambda^{\prime *}$.
We find
\begin{equation}
\mathcal{L}_{SS,\Lambda^{\prime *}}^{\rm QCD} = g_{ss}\,
\overline{\Lambda}^{\prime *}\, \eta'\, \Lambda^{\prime *},
\label{eq:fullQCDSS}
\end{equation}
and
\begin{align}
\mathcal{L}_{SS,\Lambda^{\prime *}}^{\rm Ghost} = \frac{g_{ss}}{\sqrt{2}}\, \overline{\Lambda}^{\prime *}\, \bigg \{ \, &
\sqrt{\frac{2}{3}}\, \Big ( \ 
    \tilde{K}^{- \prime}\,\tilde{\Lambda}^{+\prime *}_{p,\tilde{u}} 
  + \overline{\tilde{K}}^{0 \prime}\,\tilde{\Lambda}^{0\prime *}_{n,\tilde{d}}  \nonumber \\
&\qquad
  + \tilde{\pi}^{+ \prime}\, \tilde{\Lambda}^{-\prime *}_{\Sigma,\tilde{d}} \,
  + \tilde{\pi}^{- \prime}\, \tilde{\Lambda}^{+\prime *}_{\Sigma,\tilde{u}}  \nonumber \\
&\qquad
  + \tilde{K}^{+ \prime}\,\tilde{\Lambda}^{-\prime *}_{\Xi,\tilde{s}} 
  + \tilde{K}^{0 \prime}\,\tilde{\Lambda}^{0\prime *}_{\Xi,\tilde{s}} 
\Big ) \nonumber \\
& + \frac{1}{\sqrt{3}}\Big( \,
    \tilde{\pi}^{0 \prime} \, \tilde{\Lambda}^{0\prime *}_{\Sigma,\tilde{u}}
  - \tilde{\pi}^{0 \prime} \, \tilde{\Lambda}^{0\prime *}_{\Sigma,\tilde{d}} 
\, \Big ) \nonumber\\
& + \frac{1}{3} \, \Big ( \ 
       \tilde{\eta}'\,\tilde{\Lambda}^{0\prime *}_{\Sigma,\tilde{u}} 
  +    \tilde{\eta}'\,\tilde{\Lambda}^{0\prime *}_{\Sigma,\tilde{d}}  
  -2\, \tilde{\eta}'\,\tilde{\Lambda}^{0\prime *}_{\Sigma,\tilde{s}} 
\Big )
 \bigg \}.
\label{eq:PQQCDSS}
\end{align}
Here we have added a prime to the meson labels to remind the reader
that in full QCD these contributions enter with the meson
propagating with the mass of the flavor-singlet $\eta'$.  The symbols
serve to indicate the manner in which specific quark-flow disconnected
loops contribute to the full-QCD process.

Again, we observe the physical process proceeding through a quark-flow disconnected diagram.  Here, the
final term, indicating the ghost-$\eta'$ contributions, provides a total contribution with strength
$g_{ss}$, precisely removing the full-QCD contribution of Eq.~(\ref{eq:fullQCDSS}).  The remaining
ghost terms enter to ensure unphysical contributions from quark-flow connected diagrams do not
contribute in full QCD.  Their contributions are important and must be included when evaluating
the quark-flow disconnected contributions to baryon matrix elements.

\section{Conclusion}
\label{sec:conc}

An extension of the graded symmetry approach to partially quenched QCD
has been presented.  The extension builds on previous theory by
explicitly including flavor-singlet baryons in its construction.
The aim is to determine the strength of both the quark-flow connected
and quark-flow disconnected diagrams encountered in full-QCD
calculations of hadronic matrix elements containing flavor-singlet
baryon components.

This is particularly important in light of recent progress in the
calculation of the electromagnetic properties of the $\Lambda(1405)$
in lattice QCD \cite{Hall:2014uca}, where flavor-singlet symmetry is
known to play an important role \cite{Menadue:2011pd}.  There only the
quark-flow connected contributions are calculated.  

Because a determination of the quark-flow disconnected contributions
to baryon excited states presents formidable challenges
to the lattice QCD community, it is essential to have
an understanding of the physics missing in these contemporary lattice
QCD simulations.
An understanding of the couplings of the quark-flow disconnected
contributions to baryon matrix elements is vital to understanding QCD
and ultimately comparing with experiment.
\color{black}

We have discovered that the full-QCD processes
\begin{align*}
 &{\Lambda}^{\prime *} \to K^-\, p  \to {\Lambda}^{\prime *} \, , \quad &&{\Lambda}^{\prime *} \to \overline{K}^0\, n  \to {\Lambda}^{\prime *} \, , \nonumber \\
 &{\Lambda}^{\prime *} \to \pi^+\, \Sigma^-  \to {\Lambda}^{\prime *} \, , &&{\Lambda}^{\prime *} \to \pi^0\, \Sigma^0  \to {\Lambda}^{\prime *} \, ,   \nonumber \\
 &{\Lambda}^{\prime *} \to \pi^-\, \Sigma^+  \to {\Lambda}^{\prime *} \, , &&{\Lambda}^{\prime *} \to K^+\, \Xi^-  \to {\Lambda}^{\prime *} \, ,        \nonumber \\
 &{\Lambda}^{\prime *} \to K^0\, \Xi^0  \to {\Lambda}^{\prime *} \, \quad \mbox{ and } &&{\Lambda}^{\prime *} \to \eta\, \Lambda \to {\Lambda}^{\prime *} \, ,
\end{align*}
proceed with two thirds of the full-QCD strength in the quark-flow
disconnected diagram.  Details of the quark flavors participating in
the loops and their relative contributions are provided in
Eq.~(\ref{eq:PQQCD}) of Sec.~\ref{subsec:s2o}.

We have also considered the separation of quark-flow connected and
disconnected contributions for the flavor-singlet baryon dressings of
octet baryons in Sec.~\ref{subsec:o2s}, and flavor-singlet baryon
dressings of the the flavor-singlet baryon in Sec.~\ref{subsec:s2s}.

When combined with previous results for flavor-octet baryons, this
information enables a complete understanding of the contributions of
disconnected sea-quark loops in the matrix elements of baryons
and their excitations,} including the $\Lambda(1405)$.  This allows an
estimation of quark-flow disconnected corrections to lattice QCD
matrix elements, or an adjustment of other observables prior to making
a comparison with the quark-flow connected results of lattice QCD.  

\vspace{12pt}
\begin{acknowledgments}
We thank Tony Thomas for inspiring this study, Phiala Shanahan for
helpful discussions and Stephen Sharpe for insightful comments in the
preparation of this manuscript.  This research is supported by the
Australian Research Council through Grants No.\ DP120104627,
No.\ DP140103067 and No.\ DP150103164 (D.B.L.).
\end{acknowledgments}

\bibliography{refs}

\end{document}